\documentclass[a4paper,11pt]{article}
\usepackage{latexsym}
\usepackage{amsmath}
\usepackage{amssymb}
\usepackage{bm}
\usepackage[dvipdf,dvips]{graphics}
\title{ Two-flavor QCD phases and condensates at finite isospin chemical potential }
\author{$\mbox{Zhao Zhang}^{b,a,1}$, $\mbox{Yu-xin
Liu}^{b,a,2}$\\[5pt]
\textit{${}^a$Department of Physics, Peking University, Beijing 100871, P. R. China}\\
\textit{${}^b$CCAST(World Laboratory), P.O. Box 8730, Beijing 100080, P. R. China}
}
\date{}
\begin{document}
\maketitle \footnotetext[1]{E-mail: zhaozhang@pku.edu.cn}
\footnotetext[2]{E-mail: yxliu@pku.edu.cn}
\begin{abstract}
We study the phase structure and condensates of two-flavor QCD at
finite isospin chemical potential in the framework of a confining,
Dyson-Schwinger equation model. We find that the pion
superfluidity phase is favored at high enough isospin chemical
potential. A new gauge invariant mixed quark-gluon condensate
induced by isospin chemical potential is proposed based on
Operator Product Expansion. We investigate the sign and
 magnitude of this new condensate and show that it's an important condensate in
 QCD sum rules at finite isospin density.
\\[4pt]
\end{abstract}
\newpage
\noindent{\textbf{\Large{1. Introduction}}} \vspace{5pt}

The phase structure of QCD at non-zero temperature and baryon
chemical potential has been intensively investigated throughout
the last decade. In reality, dense baryonic matter obeys an
isospin asymmetry, i.e., in the case of two light flavors, the
densities of u and d quark are different. In order that QCD
adequately describe the isotopically asymmetric matter, such as
compact star, isospin asymmetric nucleon matter and heavy ion
collisions , usually the isospin chemical potential
$\mu_I=(\mu_u-\mu_d)$ is introduced in the theory\cite{r10, r20}.
Different approaches, such as Lattice QCD\cite{r30,r50}, chiral
perturbation theory\cite{r10,r20,r63,r65}, Ladder QCD\cite{r70},
Nambu-Jona-Lasinio type model \cite{r80,r90, r100,r110,r115}, and
random matrix model\cite{r120} has been used to explore the QCD
phase structure at finite isospin density. It has been widely
confirmed that there is a phase transition from the normal phase
to the pion superfluidity phase which is characterized by a pion
condensate $\langle{\overline{d}\gamma_5u+H.c.}\rangle$ at high
enough isospin chemical potential. It is also found that the kaon
superfluidity phase characterized by kaon condensate
$\langle{\overline{s}\gamma_5u+H.c.}\rangle$ appears at high
isospin and strangeness chemical potential in the three light
flavors case\cite{r20, r110}.
\par
The previous studies on the effects of finite isospin chemical
potential and strangeness are mostly focused on two types of
condensates $\langle{\overline{u}\gamma_5d+H.c.}\rangle$ and
$\langle{\overline{s}\gamma_5d+H.c.}\rangle$, which are order
parameters for the corresponding superfluidity phase transitions .
It is generally believed that the vacuum of QCD has complicated
structure and it is expected that all gauge invariant Lorentz
singlet local operators built of the quarks and/or gluons have
non-vanishing vacuum expectation values according to QCD sum
rules\cite{r130,r132}. For example, the well-known low-dimensional
condensates, such as quark condensate $\langle\bar{q}q\rangle$,
gluon condensate $g^2\langle{GG}\rangle$ , mixed quark gluon
 condensate $g\langle\bar{q}\sigma{G}q\rangle$ and  four quark condensate $\langle\bar{q}\Gamma_1q\bar{q}\Gamma_2q\rangle$,
 play significant roles in the hadronic studies based on QCD sum rules.
 \par
Due to the presence of the flavor mixed condensates
$\langle{\overline{u}\gamma_5d+H.c.}\rangle$ and
$\langle{\overline{s}\gamma_5d+H.c.}\rangle$ at finite isospin and
strangeness chemical potential, it is natural to expect that there
should exist other new types of flavor mixed condensates induced
by isospin chemical potential and strangeness chemical potential
according to Operator Product Expansion(OPE). Besides the pion
condensate and kaon condensate, the possible low-dimensional
flavor mixed condensates are  mixed quark-gluon condensate
 $g\langle\overline{d}\gamma_5\sigma{G}u+H.c.\rangle$ induced by isospin density, and
$g\langle\overline{s}\gamma_5\sigma{G}u+H.c.\rangle$ induced by strangeness density
(for convenience, we call the former pion mixed quark-gluon condensate and the later
kaon mixed quark-gluon condensate). In addition,  new forms of four quark condensates,
such as $\langle\bar{q}\gamma_5\tau_i\Gamma_1q\bar{q}\Gamma_2q\rangle$, may also appear in OPE.
 We expect these  induced low-dimensional condensates also play important
roles on the hadronic physical observables in the framework of QCD
sum rules.
\par
It is well-known that, in the chiral limit, both chiral condensate
and mixed quark-gluon condensate are ideal order parameters for
the chiral phase transition of QCD. Similarly, pion mixed
quark-gluon condensate and kaon mixed quark-gluon condensate can
play the roles of order parameters for the pion superfluidity
phase transition and the kaon superfluidity phase transition at
finite isospin chemical potential and strangeness chemical
potential, respectively.
 Though both pion condensate and pion mixed quark-gluon
condensate can be used as order parameters to describe the pion
superfluidity phase transition, they reflect different aspects of
the non-perturbative structure of the ground state: the former
reflects the correlation between different flavors with
color-singlet component, while the later reflects the correlation
between different flavors with color-octet components. Therefore,
pion mixed quark-gluon condensate will give new and important
information on the pion superfluidity phase transition. The same
thing is true for kaon condensate and kaon mixed quark-gluon
condensate.
\par
Therefore, it is interesting to investigate the thermal and dense
properties of these new types of low-dimensional condensates and
their effects on the physical hadronic observables. Since the
Global Color Model (GCM)\cite{r140,r142,r144,r146} is an effective
quark and gluon fields theory and  has been successfully used to
investigate the property of the traditional mixed quark-gluon
condensate\cite{r150, r152} and other QCD condensates\cite{r154},
we will adopt this model to explore the thermal and dense
properties of above induced mixed condensates. In this paper, we
only consider the possible pion superfluidity phase transition and
pion mixed quark-gluon condensate.
\par

\vspace{5pt} \noindent{\textbf{\Large{2. Mean field theory of GCM
at finite isospin chemical potential}}}
 \vspace{5pt}

In Euclidean metric, with ${\{\gamma_\mu,\gamma_\nu\}}=
2\delta_{\mu\nu}$ and $\gamma_\mu^{+}=\gamma_\mu$, the generating functional of
GCM with quark and gluon degrees is
\begin{equation}
Z[J,\overline{\eta},\eta]={\int}D\overline{q}D{q}DAexp({-S_{GCM}[\overline{q},q,A_\mu^a]+\overline{\eta}q+\overline{q}\eta+J_\mu^{a}A_\mu^a})
\end{equation}
with the action
\begin{equation}\label{ACT}
S_{GCM}[\overline{q},q,A_\mu^a]=\int(\overline{q}(\gamma\cdot\partial+M-igA_\mu^a\frac{\lambda^a}{2}\gamma_\mu)q+\frac{1}{2}A_\mu^aD^{-1}_{\mu\nu}(i\partial)A_\nu^a).
\end{equation}
The essence of GCM is that it models the QCD local gluonic action
$\int{F^a_{\mu\nu}}{F^a_{\mu\nu}}$ which has local color symmetry,
by a highly nonlocal action which has a global color symmetry. The
main aspects of GCM have been reviewed in {\cite{r142,r144,r146}}.

\par
Integrating over the gluon degrees, the partition function of the
GCM with two quark flavors at finite baryon and isospin chemical
potential(we only consider the case with temperature $T=0$ in this
paper) has the form
\begin{eqnarray}
Z(\mu,\mu_I)&=&\int{D\bar{q}(x)Dq(x)}\text{exp}
[-\int_x\overline{q}(x)[\gamma_\mu\partial_\mu+M-\mu\gamma_4-\delta\mu\tau_3\gamma_4]q(x)\nonumber\\
&&-\frac{1}{2}\int_x\int_yj^a_{\mu}(x)g^2D_{\mu\nu}(x-y;\mu,\mu_I)j^a_\nu(y)]\label{PTF},
\end{eqnarray}
where $M=diag(m_u,m_d)$, $\int_x=\int{d^4{x}}$ and
$j^a_\mu(x)=\bar{q}(x)\gamma_\mu\frac{\lambda^a_C}{2}q(x)$. In
Eq.\eqref{PTF}, $\tau_i(i=1,2,3)$ are the Pauli matrixes in flavor
place,
 $\mu\equiv\mu_B/3$ is the chemical potential associated with baryon number and the
 quantity $\delta\mu$ is a half of the isospin
chemical potential, i.e. $\delta\mu=\mu_I/2$ (In this paper, we only
consider $\mu_I>0$). The effective gluon propagator
$g^2D_{\mu\nu}(x-y;\mu,\mu_I)$ is generally a
$(\mu,\mu_I)$-dependent function, which is parameterized to model
the low energy dynamics of QCD.
\par
In this study, we will take $m_u=m_d=m$. Evidently, the above
Lagrangian is invariant under the baryon $U_B(1)$ symmetry and the
parity symmetry transformation P. In the case with $\mu_I\neq{0}$,
the traditional isospin $SU_I(2)$ symmetry is reduced to
$U_{I_3}(1)$ symmetry. Usually, the quark condensate
$\langle\bar{q}q\rangle$ is responsible for the chiral symmetry
breaking of the ground state and does not spoil the parity and
isotopical symmetry, while the nonzero pion condensate
$\langle\bar{q}\gamma_5\tau_1q\rangle$ breaks  both the parity and
isotopical symmetry of the ground state.
\par
Within the GCM formalism, the ground state of QCD is defined by the
 saddle point of the action and
 the quark gap equation at the mean field level is determined
by the rainbow truncated quark Dyson-Schwinger equation (DSE) (The
application of the DSE model to finite temperature and density is
reviewed in \cite{r160})
\begin{eqnarray}\label{DSE}
\Sigma(p)=\frac{4}{3}\int\frac{d^4q}{(2\pi)^4}g^2D_{\mu\nu}(p-q)\gamma_{\mu}S(q)\gamma_\nu.
\end{eqnarray}
At finite $(\mu, \mu_I)$ with u, d quarks, the inverse of quark
propagator can be written in the form
\begin{equation}
S^{-1}(p,\mu,\mu_I)=S_0^{-1}(p,\mu,\mu_I)+\left(\begin{array}{cc}
    \Sigma_{uu}(p,\mu,\mu_I)&\Sigma_{ud}(p,\mu,\mu_I)\\
    \Sigma_{du}(p,\mu,\mu_I)&\Sigma_{dd}(p,\mu,\mu_I)
    \end{array}\right),
\end{equation}
where
\begin{equation}
S_0^{-1}(p,\mu,\mu_I)=\left(\begin{array}{cc}
    i\vec{\gamma}\cdot\vec{p}+i\gamma_4w_u+m&\\
    &i\vec{\gamma}\cdot\vec{p}+i\gamma_4w_d+m
    \end{array}\right),
\end{equation}
with $w_u=(p_4+i\mu+i\delta\mu)$, $w_d=(p_4+i\mu-i\delta\mu)$,
\begin{eqnarray}
\Sigma_{aa}(p,\mu,\mu_I)&=&i\vec{\gamma}\cdot\vec{p}A_a(\vec{p},w_u,w_d)+i\gamma_4w_aB_a(\vec{p},w_u,w_d)+C_a(\vec{p},w_u,w_d),\\
\Sigma_{ud}(p,\mu,\mu_I)&=&\Sigma_{du}(p,\mu,\mu_I)=i\gamma_5D(\vec{p},w_u,w_d),
\end{eqnarray}
and $A_a, B_a, C_a, D$ are momentum-dependent scalar functions.
Nonzero $C_a$ and $D$ are responsible for the dynamical chiral
symmetry breaking and isotopical symmetry breaking, respectively.

\par
 Note that the possible
diquark condensation is not considered here and only single
Lorentz structure is concerned in $\Sigma_{ud}$ and $\Sigma_{du}$.
The four matrix elements of the momentum dependent quark
propagator
\begin{equation}
S(p,\mu,\mu_I)=\left(\begin{array}{cc}
    S_{uu}(p,\mu,\mu_I)&S_{ud}(p,\mu,\mu_I)\\
    S_{du}(p,\mu,\mu_I)&S_{dd}(p,\mu,\mu_I)
    \end{array}\right)
\end{equation}
take the form
\begin{gather}
\begin{split}
S_{uu}&=[(X_dC_u+D^2C_d)-i\vec{\gamma}\cdot\vec{p}(X_dA_u+D^2A_d)-i\gamma_4(X_dw_uB_u+D^2w_dB_d)]/H,\\
S_{dd}&=[(X_uC_d+D^2C_u)-i\vec{\gamma}\cdot\vec{p}(X_uA_d+D^2A_u)-i\gamma_4(X_uw_dB_d+D^2w_uB_u)]/H,\\
S_{ud}&=-iD[\gamma_5Y-i\gamma_5\vec{\gamma}\cdot\vec{p}S_{ud}^{\gamma_5\vec{\gamma}}
-i\gamma_5\gamma_4S_{ud}^{\gamma_5\gamma_4}+\gamma_5\vec{\gamma}\cdot\vec{p}\gamma_4S_{ud}^{\gamma_5\vec{\gamma}\gamma_4}]/H,\\
S_{du}&=-iD[\gamma_5Y-i\gamma_5\vec{\gamma}\cdot\vec{p}S_{du}^{\gamma_5\vec{\gamma}}
-i\gamma_5\gamma_4S_{du}^{\gamma_5\gamma_4}-\gamma_5\vec{\gamma}\cdot\vec{p}\gamma_4S_{du}^{\gamma_5\vec{\gamma}\gamma_4}]/H,
\end{split}
\end{gather}
with
\begin{eqnarray}\label{EXTR}
\begin{aligned}
S_{ud}^{\gamma_5\vec{\gamma}}&=-S_{du}^{\gamma_5\vec{\gamma}}=A_uC_d-C_uA_d,\\
S_{ud}^{\gamma_5\gamma_4}&=-S_{du}^{\gamma_5\gamma_4}=w_uB_uC_d-w_dB_dC_u,\\
S_{ud}^{\gamma_5\vec{\gamma}\gamma_4}&=-S_{du}^{\gamma_5\vec{\gamma}\gamma_4}=A_uw_dB_d-A_dw_uB_u,
\end{aligned}
\end{eqnarray}
and
\begin{gather}
X_a=A_a^2{\vec{p}}^2+B_a^2w_a^2+C_a^2, \quad Y=C_uC_d+A_uA_d\vec{p}^2+w_uw_dB_uB_d+D^2,\\
H=X_uX_d+D^4+2D^2(C_uC_d+A_uA_d\vec{p}^2+w_uw_dB_uB_d).
\end{gather}
With above decomposition, the gap equation can be expressed as
\begin{equation}\label{GAP}
\Sigma_{ij}(p)=\frac{4}{3}\int\frac{d^4q}{(2\pi)^4}g^2D_{\mu\nu}(p-q)\gamma_{\mu}S_{ij}(q)\gamma_\nu.
\end{equation}
\par
Since each term of Eq.\eqref{EXTR} is nonzero at finite $\mu_I$,
more Lorentz structures with new scalar
 functions should also be considered in $\Sigma_{ud}$ and $\Sigma_{du}$
  to guarantee the self-consistent
 treatment of the gap equation. However, introducing more Lorentz structures will complicate the
resolving of the gap equations. Just as the Lorentz tensor
structure is not concerned in $\Sigma_{aa}$ at the traditional
treatment of DSE at finite $(T,\mu)$\cite{r160},
 we suppose that $i\gamma_5D$ is the leading order term of $\Sigma_{ud}$ and $\Sigma_{du}$ and
 other Lorentz structures has small impact on the determination of quark self energy.
 Because there are no structures $\gamma_5\vec{\gamma}$, $\gamma_5\gamma_4$ and $\gamma_5\vec{\gamma}\gamma_4$ in
 $\Sigma_{ud}$ and $\Sigma_{du}$, the corresponding structures associated with Eq.\eqref{EXTR} in
  $S_{ud}$ and $S_{du}$ are
 ignored in the following. At least, this is a good approximation in the case with small $\mu_I$.
\par
Due to the phenomenological nature of this effective theory, for
simplicity, the Feynman-like gauge
$g^2D_{\mu\nu}(p-q)=\delta_{\mu\nu}g^2D(p-q)$ was adopted in our
calculation. With above approximation, the gap equation
\eqref{GAP} is reduced to seven coupled integral equations, which
is still complicated to solve. To get a qualitative understanding
of the phase diagram and the structure of the ground state at
finite $\mu_I$, a pedagogical model first introduced by Munczek
and Nemirovsky \cite{r170} for the modelling  of confinement in
QCD is favored in this study. Munczek-Nemirovsky(MN) model has
been extensively used to explore the properties of strong QCD both
at zero $(T,\mu)$ and nonzero $(T,\mu)$\cite{r180,r152}, which can
always give qualitatively consistent results with the more
sophisticated models. The effective gluon propagator of MN model
takes the form
\begin{equation}\label{MN}
g^2D_{\mu\nu}(p-q)=\delta_{\mu\nu}\frac{3}{16}(2\pi)^4{\eta^2}\delta^4(p-q),
\end{equation}
with the single parameter $\eta$ determined by $\pi$ and $\rho$
masses in vacuum. The scale parameter $\eta$ has relation with the
string tension of QCD, and in the more real world, it should been
the function of $T$ and $\mu$. Using Eq.\eqref{MN}, the complete
expressions of Eq.\eqref{GAP} are simplified
 as seven-coupled algebraic equations
\begin{gather}
\label{GAPA}(A_u-1)=\frac{1}{2}\eta^2[{X_dA_u+D^2A_d}]/H,  \quad  (A_d-1)=\frac{1}{2}\eta^2[{X_uA_d+D^2A_u}]/H,\\
\label{GAPB}(B_u-1)=\frac{1}{2}\eta^2[X_dB_u+\frac{w_{d}}{w_{u}}D^2B_d]/H, \quad (B_d-1)=\frac{1}{2}\eta^2[X_uB_d+\frac{w_{u}}{w_{d}}D^2B_u]/H,\\
\label{GAPC}C_u-m=\eta^2[{X_dC_u+D^2C_d}]/{H}, \quad C_d-m=\eta^2[{X_uC_d+D^2C_u}]/{H},\\
\label{GAPD}D=\eta^2D[{C_uC_d+A_uA_d\vec{p}^2+w_{u}w_{d}B_uB_d+D^2}]/{H}.
\end{gather}
Eq.\eqref{GAPD} illustrates that there are two distinctive
solutions to $D$: one characterized by $D\equiv{0}$, which
describes the normal phase;  the alternative, characterized by
$D\neq{0}$, which describes the pion superfluidity phase. The
phase with small free energy is favored in nature.
\par
\vspace{5pt} \noindent{\textbf{\Large{3. Thermal Potential and
Condensates }}} \vspace{5pt}\par In GCM/DSE formalism, whether the
normal phase or the pion superfluidity phase is stable is
determined by evaluating the $(\mu,\mu_I)$-dependent pressure
difference
\begin{equation}
\delta{P}(\mu,\mu_I)=P[\mu,\mu_I,S[D\neq{0}]]-P[\mu,\mu_I,S[D={0}]],
\end{equation}
where the pressure is calculated by using a steepest-descent
approximation\cite{r190}:
\begin{equation}
P[T,\mu,\mu_I,S]=-\Omega[S]=\frac{1}{{\beta}V}TrLn[\beta{S^{-1}}]-\frac{1}{{\beta}V}TrLn[\Sigma{S}].
\end{equation}
Using the technique
\begin{equation}
Det\left(\begin{array}{cc}
    A & B\\
    C & D\end{array}
 \right)=Det(A)Det(B)Det(C)Det(C^{-1}DB^{-1}-A^{-1}),
\end{equation}
the pressure at finite $(\mu,\mu_I)$ can be expressed as
\begin{equation}\label{ThermalP}
\begin{split}
P[S]=&\int_{p}2Ln[H]+2N_c\int_{p}\Big[D^2[\vec{p}^2(A_u+A_d)+w_{u}w_{d}(B_u+B_d)]+X_u[\vec{p}^2A_d+w_{d}^2B_d]\\
&+X_d[\vec{p}^2A_u+w_{d}^2B_u]+m[X_dC_u+X_uC_d+D^2(C_u+C_d)]\Big]/H,
\end{split}
\end{equation}
where $\int_{p}=\int{\frac{d^4{p}}{(2\pi)^4}}$. Note that a
constant term has been ignored in Eq.\eqref{ThermalP}. Though the
pressure (or thermal potential) calculated through
Eq.\eqref{ThermalP} is ultraviolet divergent, the pressure
difference or the ``bag constant" $\delta{P}$ \cite{r140} is
finite.
\par
From the GCM generating functional, it is straightforward to
calculate the vacuum expectation value(VEV) of any quark operator
with the forms
\begin{equation}\label{qop}
\mathcal{O}_n\equiv(\overline{q}_{j_1}\Lambda^{(1)}_{j_1i_1}q_{i_1})
(\overline{q}_{j_2}\Lambda^{(2)}_{j_2i_2}q_{i_2})\cdots
(\overline{q}_{j_n}\Lambda^{(n)}_{j_ni_n}q_{i_n}) ,
\end{equation}
in the mean field vacuum. Here $\Lambda^{(i)}$ stands for an
operator in Dirac, flavor, and color space.  The VEV of the
operator $\mathcal{O}_n$ has the form\cite{r200}
\begin{eqnarray}\label{qvev}
\langle\mathcal{O}_n\rangle=(-1)^n\sum_p(-)^p[\Lambda^{(1)}_{j_1i_1}\cdots
\Lambda^{(n)}_{j_ni_n}S_{i_1j_{p(1)}}{\cdots}S_{i_nj_{p(n)}}],
\end{eqnarray}
where $p$ stands for a permutation of the $n$ indices. Based on
formula \eqref{qvev}, the low-dimensional condensates, such as
chiral condensate and pion condensate can been expressed as
\begin{eqnarray}
\langle\bar{u}u\rangle=-Tr_{D,C}[S_{uu}(x,x)-\sigma_{UV}(x,x)]&=&-N_c\int_{p}Tr_D[S_{uu}(p)-\sigma_{UV}(p)],\label{uucond}\\
\langle\bar{d}d\rangle=-Tr_{D,C}[S_{dd}(x,x)-\sigma_{UV}(x,x)]&=&-N_c\int_{p}Tr_D[S_{dd}(p)-\sigma_{UV}(p)],\label{ddcond}\\
\langle\bar{q}i\gamma_5\tau_1q\rangle=-Tr_{D,C,F}[i\gamma_5\tau_1S(x,x)]
&=&-N_c\int_{p}Tr_D[i\gamma_5S_{ud}(p)+i\gamma_5S_{du}(p)].
\end{eqnarray}
To get a convergent condensate integral, a subtracting term
$\sigma_{UV}(p)$ which simulate the ultraviolet behavior of the
quark propagator is introduced in the definition of the quark
condensate. In the case with nonzero current quark mass and zero
chemical potential, for MN model one has, with $s=p^2$,

\begin{equation}
A(s)=B(s)=1+\frac{1}{2s}, \quad
C(s)=m(1+\frac{2}{s}),
\end{equation} with corrections of high
order in ($1/s$). From these approximate expressions one can
construct
\begin{equation}
\sigma_{UV}(s)=\frac{C(s)}{sA(s)^2+C(s)^2}.
\end{equation}
Since $\sigma_{UV}(s)\rightarrow{m/s}$ as $s\rightarrow{\infty}$
and $\sigma_{UV}(s)\rightarrow{0}$ as $s\rightarrow{0}$, this
prescription will provide an absolutely convergent result with no
need for a cutoff. The above definition can be generalized to the
case with nonzero quark chemical potential. There is no need to
introduce a subtracting term in the definition of the pion
condensate since  $D(p)$ keeps zero in the large momentum region
in MN model. The formula for evaluating the 6-dimensional
four-quark condensates can also be directly derived from
Eq.\eqref{qvev}, which is consistent with the vacuum saturation
approximation at zero quark chemical potential. At finite isospin
density, the new type four-quark condensates,  such as
$\langle\bar{q}i\gamma_5\tau_1q\bar{q}q\rangle$, will appear in
OPE.
 It is easily to prove that $\langle\bar{q}i\gamma_5\tau_1q\bar{q}q\rangle\sim\langle\bar{q}i\gamma_5\tau_1q\rangle\langle\bar{q}q\rangle$
with the approximation that only $\gamma_5$ structure is hold in
$\Sigma_{ud(du)}$ and $S_{ud(du)}$,  with
$\langle\bar{q}q\rangle=\langle\bar{u}u+\bar{d}d\rangle$.
\par
Since the functional integration over the gluon field $A^a_\mu$ is
quadratic in the framework of GCM, one can perform the integration
over gluon field analytically. Using the technique introduced by
Meissner\cite{r150}, through the following integral formulae
\begin{equation}\label{gvev}
\begin{split}
\int\mathcal{D}Ae^{-\frac{1}{2}AD^{-1}A+jA}
&={e}^{\frac{1}{2}jDj},\\
\int\mathcal{D}AAe^{-\frac{1}{2}AD^{-1}A+jA}
&=(jD){e}^{\frac{1}{2}jDj},\\
\int\mathcal{D}AA^2e^{-\frac{1}{2}AD^{-1}A+jA}
&=[D+{(jD)}^2]{e}^{\frac{1}{2}jDj},\\
&\cdots.
\end{split}
\end{equation}
the gluon fields vacuum average can be replaced by the quark
current  $j_{\mu}^a$ with the effective gluon propagator $D(x-y)$.
At the mean field level, according to Eq.\eqref{qvev}, one can in
principle obtain the VEVs for any gluon fields. This technique
provides a feasible way to calculate the VEVs of operators with
low-dimensional gluon fields such as the traditional mixed
quark-gluon condensate and the isospin density induced pion mixed
quark-gluon condensate. Since the number of terms produced by
Eq.\eqref{qvev} will increased rapidly with the number of gluonic
fields, this technique is not suitable for the evaluation of the
VEV of the operator involving high powers of gluonic field $A$.
For instance for the gluon condensate $\langle{GG}\rangle$, which
contains a $A^4$ term, the calculation gets already rather
involved.
\par
Applying the method described above, we obtain the expression
\begin{equation}\label{pmixd}
\begin{split}
g\langle\bar{q}i\gamma_5\tau_1\sigma_{\mu\nu}G_{\mu\nu}^a\frac{\lambda_c^a}{2}q\rangle=&-2iN_c\int_y\frac{4}{3}
[\partial_{\mu}^xg^2D(y-x)]Tr_{D,F}[S(y-x)i\gamma_5\tau_1\sigma_{\mu\nu}S(x-y)\gamma_{\nu}]\\
&+4iN_c\int_{y}\int_{z}g^2D(y-x)g^2D(z-x)Tr_{D,F}[S(z-x)i\gamma_5\tau_1\sigma_{\mu\nu}\\
&{\times}S(x-y)\gamma_{\mu}S(y-z)\gamma_{\nu}].
\end{split}
\end{equation}
The similar expression for evaluating the traditional quark-gluon
condensate
$g\langle\bar{q}\sigma_{\mu\nu}G_{\mu\nu}^a\frac{\lambda_c^a}{2}q\rangle$
can be got by replacing the structure
$i\gamma_5\tau_1\sigma_{\mu\nu}$ in \eqref{pmixd} with
$\sigma_{\mu\nu}$.
\par
Using the Eq.\eqref{GAP} and the formulae
\begin{equation}
Tr_c\Big[\frac{\lambda_c^a}{2}\frac{\lambda_c^b}{2}\frac{\lambda_c^c}{2}-\frac{\lambda_c^a}{2}\frac{\lambda_c^c}{2}\frac{\lambda_c^b}{2}\Big]
=\frac{i}{2}f^{abc}, \quad  f^{abc}f^{abc}=N_c\delta^{aa},
\end{equation}
the expression for the pion mixed quark-gluon condensate can be
simplified as
\begin{equation}\label{exppmc}
g\langle\bar{q}i\gamma_5\tau_1\sigma_{\mu\nu}G_{\mu\nu}^a\frac{\lambda_c^a}{2}q\rangle=I_1+I_2+I_3+I_4+I_5+I_6+I_7,
\end{equation}
where
\begin{gather}
I_1=-72\int_{p}\frac{D}{H}Y\Big[(A_u+A_d-2)]\vec{p}^2+(B_u-1)w_u^2+(B_d-1)w_d^2\Big],\\
I_2=36\int_{p}\frac{D}{H}\Big[X_dA_u+D^2A_d+X_uA_d+D^2A_u\Big]\vec{p}^2,\\
I_3=36\int_{p}\frac{D}{H}\Big[X_dB_uw_u^2+D^2w_uw_dB_d+X_uB_dw_d^2+D^2w_uw_dB_u\Big],\\
I_4=\frac{81}{2}\int_{p}\frac{D}{H}Y\Big[D^2-(C_u-m)(C_d-m)\Big],\\
I_5=\frac{81}{2}\int_{p}\frac{D}{H}\Big[(C_u-m)(X_dC_u+D^2C_d)+(C_d-m)(X_uC_d+D^2C_u)\Big],\\
I_6=\frac{81}{2}\int_{p}\frac{D}{H}\Big[(A_u-1)(X_dA_u+D^2A_d)+(A_d-1)(X_uA_d+D^2A_u)\Big]\vec{p}^2,\\
I_7=\frac{81}{2}\int_{p}\frac{D}{H}\Big[(B_u-1)(X_dB_uw_u^2+D^2B_dw_uw_d)+(B_d-1)(X_uB_dw_d^2+D^2B_uw_uw_d)\Big].
\end{gather}
Note that the scalar functions $A_a, B_a, C_a, D$ are all momentum
dependent in the DSEs formalism. The above integral expressions
explicitly show that the pion mixed quark-gluon condensate is a
order parameter for the pion superfluidity phase transition. By
the way, it should be mentioned that the expression in
Eq.\eqref{exppmc} for pion mixed quark-gluon condensate is only
valid in Feynman-like gauge.
\par
\vspace{5pt}
\noindent{\textbf{\Large{4. Numerical Results and
Discussions }}} \vspace{5pt}
\par
To get a qualitative understanding of the finite $\mu_I$ effects
on the ground state, the numerical study below are all based on
the MN model. Due to the defect of the supposed Lorentz structure
of quark self-energy,
 the isospin chemical potential concerned below is limited within the range $|\mu_I/2|<0.2(\text{GeV})$.
 In this paper, for simplicity, we only concern the case with zero temperature and baryon chemical potential.

\par
\vspace{5pt}
\noindent{\textbf{\large{A. The critical isospin
chemical potential for pion condensate}}}
\par
The effective field theory arguments \cite{r10} indicate that
critical isospin chemical potential $\mu_I^c$ for the pion
superfluidity phase is exactly the vacuum pion mass $m_\pi$ at
$T=\mu_B=0$. Whether the MN model can algebraically reproduce this
result within above formalism is discussed below.
\par
For the solution with $D{\neq}0$ , the gap equation (\ref{GAPD})
is reduced to
\begin{equation}\label{GapD1}
H=\eta^2[{C_uC_d+A_uA_d\vec{p}^2+B_uB_dw_{u}w_{d}+D^2}].
\end{equation}
In the neighborhood of ${\mu_I}_c$, one can probably neglect the
$D^2$ and $D^4$ in above expression. According to gap equations
(\ref{GAPA}) and (\ref{GAPB}), $A_{u(d)}$ and $B_{u(d)}$ are
identical for $D=0$. Therefore, the gap equation (\ref{GapD1}) can
be further reduced to
\begin{equation}\label{GapD2}
X(p+\frac{P'}{2})X(p-\frac{P'}{2})=\eta^2\bigg[{C(p+\frac{P'}{2})C(p-\frac{P'}{2})+A(p+\frac{P'}{2})A(p-\frac{P'}{2})(p+\frac{P'}{2})\cdot{(p-\frac{P'}{2})}}\bigg],
\end{equation}
where $P'=(\vec{0},i\mu_I^c)$ and the flavor subscript has been
ignored. Within the feynman-like gauge and only considering the
$\gamma_5$ structure, the Bethe-Salpeter equation (BSE)  for the
vacuum pseudoscalar amplitude $\Gamma_{\pi}^j(p,P)$ in GCM takes the
form \cite{r142}
\begin{equation}\label{BSE}
\Gamma_{\pi_j}(p,P)=-\frac{2}{9}\int{\frac{d^4q}{(2\pi)^4}}D(p-q)Tr_{D,C,F}[i\gamma_5{\tau_j}^+S(q+\frac{P}{2})i\gamma_5{\tau_j}S(q-\frac{P}{2})]\Gamma_{\pi_j}(q,P),
\end{equation}
where $P^2=-{m_\pi^2}$. Using $Tr_{F}[\tau_j^+\tau_j]=2$, the BSE
(\ref{BSE}) is greatly simplified in the MN model

\begin{equation}
X(p+\frac{P}{2})X(p-\frac{P}{2})=\eta^2\bigg[A(p+\frac{P}{2})A(p-\frac{P}{2})(p+\frac{P}{2})\cdot(p-\frac{P}{2})+C(p+\frac{P}{2})C(p-\frac{P}{2})\bigg],
\end{equation}
which has the same form as the gap equation (\ref{GapD2}).
Therefore, if the BSE (\ref{BSE}) can produce the vacuum pion mass,
we algebraically prove $-P'^2={m_\pi^2}$ and get the conclusion
$\mu_I^c=m_\pi$ at $T=\mu=0$ .
\par
However, the above proof is only true for the chiral limit case
since the MN model doesn't support the pion bound state beyond the
chiral limit if only the $\gamma_5$ structure is considered in the
pseudoscalar meson Bethe-Salpeter amplitude \cite{r170}. Therefore,
to produce the result $\mu_I^c=m_\pi$ beyond chiral limit in the MN
model, other Dirac amplitudes beyond $\gamma_5D$ should also be
included in the off-diagonal term of the inverse quark propagator,
which will make solving the gap equation more involved.
\par
 Note that for
other improved effective gluon propagators such as those used in
\cite{r150,r205,r210}, the pion Bethe-Salpeter amplitude with only
$\gamma_5$ structure is a good approximation to obtain the vacuum
pion mass. In principle, using these improved effective gluon
propagator to explore the pion superfluidity within the DSE
formalism should give more quantitatively reasonable results in
contrast with the simple MN model. However, the set of seven coupled
algebraic gap equations \eqref{GAPA}-\eqref{GAPD} will be replaced
by a set of seven coupled integral equations which leads to the
numerical calculation more difficult.

\par
For simplicity
 and to get a qualitatively understanding of the pion superfluidity
within the DSE formalism, we still use the MN model in the following
and only contains the $\gamma_5$ structure in the off-diagonal term
of the inverse quark propagator.
\par
\vspace{5pt}
\noindent{\textbf{\large{B. The Chiral limit}}}

\begin{figure}[ht]
\centering
 {\includegraphics{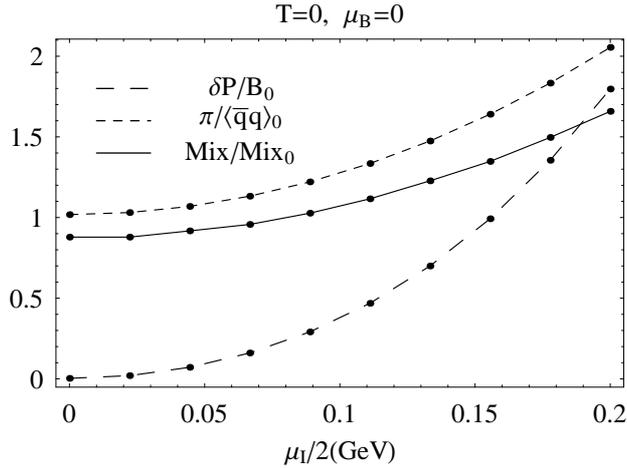}}
 \caption{The pressure difference $\delta{P}$, pion condensate $\pi$ and  pion mixed quark-gluon condensate \text{Mix} in the
   chiral limit. $\text{B}_0=(0.1\eta)^4$, $\langle\bar{q}q\rangle_0=\langle\bar{u}u+\bar{d}d\rangle_0$ and $\text{Mix}_0=(310\text{MeV})^5$ are the bag constant,
   chiral condensate and mixed quark-gluon condensate of the vacuum obtained from the MN model with $m=0$ and $\eta=1.06\text{GeV}$, respectively.}
 \label{F1}
\end{figure}
\par
In the chiral limit, there are four possible solutions according to
the gap equations \eqref{GAPA}-\eqref{GAPD}:
\begin{equation}
\begin{split}
&C_u=C_d=0, D=0; \qquad C_u\neq{0}, C_d\neq{0}, D=0;\\
&C_u=C_d=0, D\neq{0}; \qquad  C_u\neq{0}, C_d\neq{0}, D\neq{0},
\end{split}
\end{equation}
which characterize four possible phases of QCD at finite isospin
density, respectively. However, the solution with both nonzero
quark condensate and nonzero pion condensate is not found in our
numerical study. By comparing the corresponding free energies of
the former three possible phases,
 it is found that the pion superfluidity phase is favored in the chiral limit at nonzero isospin chemical potential.
 It seems that
 this is a universal result which has been confirmed by many former studies\cite{r10,r20}. In this case,
 chiral symmetry is not broken and both quark condensate and mixed quark-gluon condensate disappear
 in OPE. In contrast with the vanishing of above chiral condensates,
 new condensates induced by isospin density,
 such as pion condensate and pion mixed quark-gluon condensate
 appear. Note that the new four-quark condensates,  such as
 $\langle\bar{q}i\gamma_5\tau_1q\bar{q}q\rangle$,
also vanish in this case because these condensates factorize at
mean field level within our formalism with the approximation used
in Sec. 2.
\begin{figure}[ht]
\centering
  {\includegraphics{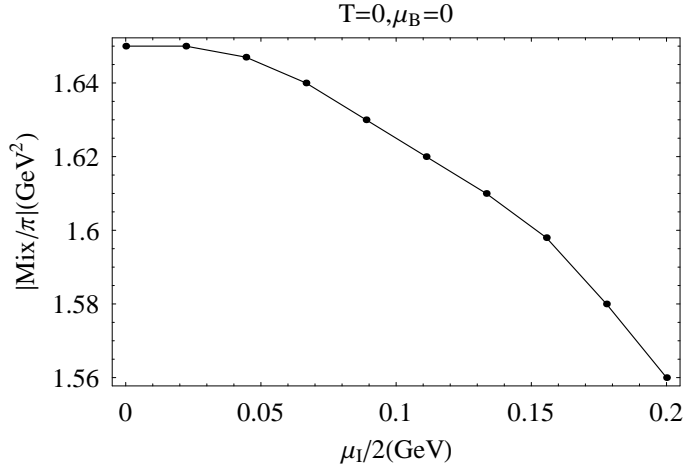}}
  \caption{The isospin chemical potential dependence of the ratio between pion mixed quark-gluon condensate
  and pion condensate from MN model with $m=0$ and $\eta=1.06\text{GeV}$.}
    \label{F2}
\end{figure}
 \par
 The $\mu_I$ dependence of the pressure difference between the pion superfluid phase and the normal phase,
  pion condensate and pion mixed quark-gluon condensate
 is shown in Fig.1. For the case with $\mu_I=0$, the zero pressure difference of
$\delta{P}$ suggests that the nonzero pion condensate and nonzero
quark condensate correspond to equivalent but distinct vacua,
which is guaranteed by the chiral symmetry(A small quark mass will
destabilize the superfluidity phase). In contrast with this event,
for the whole domain of nonzero $\mu_I$ concerned, the positive
pressure difference $\delta{P}$  suggests the pion superfluidity
phase is the stable ground state in the chiral limit for
two-flavor QCD. Fig.1 shows that the magnitudes of both induced
condensates  and the pressure difference are monotonically
increasing functions of $\mu_I$.
\par
 Fig.2 shows the $\mu_I$-dependent behavior of the ratio of pion mixed quark-gluon condensate
 to pion condensate from MN model. In Fig.2, The ratio ranges from $1.56(\text{GeV})^2$ to $1.66(\text{GeV})^2$,
 which suggests that the induced mixed quark-gluon condensate has the same magnitude of the traditional mixed condensate in
 the concerning domain of $\mu_I$(in MN model, the ratio of the mixed quark-gluon condensate and the chiral
 condensate is 1.92 in the vacuum\cite{r152}). Note that numerical study suggests that
 the pion mixed quark-gluon condensate defined in Eq.\eqref{pmixd} is
 positive, which is consistent with the sign of the traditional mixed quark-gluon
 condensate.
 \par
\vspace{5pt}
\noindent{\textbf{\large{C. Finite current mass}}}

\begin{figure}[ht]
\centering {\includegraphics{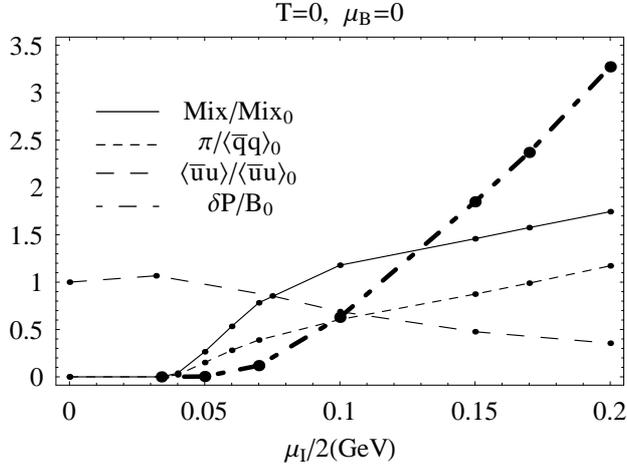}}
   \caption{The pressure difference $\delta{P}$ ( plotted only in the region $\mu_I\geq\mu_I^C$  ), quark condensate $\langle\bar{u}u\rangle$, pion condensate $\pi$ and  pion mixed quark-gluon condensate Mix
  obtained from MN model with $m=12\text{MeV}$ and $\eta=1.06\text{GeV}$.}
    \label{F3}
\end{figure}
\par
With finite current quark mass $m=12(\text{MeV})$\cite{r170},
apparently, only two types of solutions exist: the normal phase with
nonzero quark condensate and zero pion condensate and the pion
superfluidity phase with nonzero pion condensate and nonzero quark
condensate. In contrast to the chiral limit, for finite current
mass, there is no solution with zero quark condensate to the gap
equations due to the explicit chiral symmetry broken term in the
lagrangian.
\par
According to the numerical study, in the small isospin chemical
potential region $\mu_I<64\text{MeV}$, only the normal phase
solution exists.  For the region $\mu_I>64\text{MeV}$, there also
exist the solution corresponding the pion superfluidity phase beside
  the normal phase solution. The stable ground state
in the region $\mu_I>64\text{MeV}$ is again determined by the
difference of the pressure $\delta{P}$ in Eq.\eqref{ThermalP}.

\begin{figure}[ht]
 \centering {\includegraphics{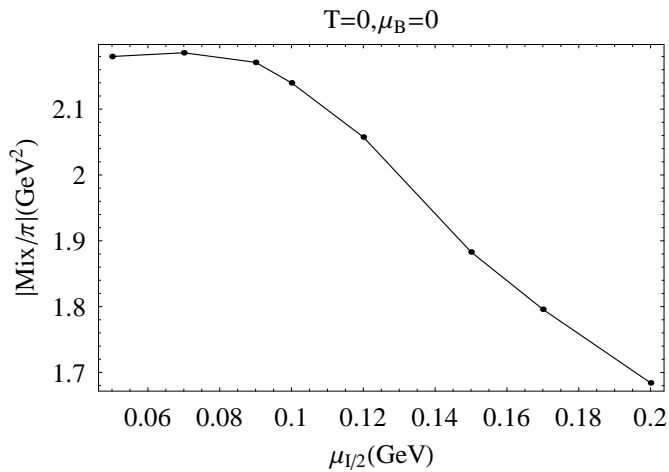}}
  \caption{The $\mu_I$-dependence of the ratio between pion mixed quark-gluon condensate
  and pion condensate obtained from MN model with $m=12\text{MeV}$ and $\eta=1.06\text{GeV}$.}
    \label{F4}
\end{figure}

\par
It is shown in Fig.3 that at the point $\mu_I=64 \text{MeV}$, the
scaled pressure difference of two solutions is close to zero and for
the region $\mu_I>64 \text{MeV}$, the pressure difference is
positive and monotonically increase with the increase isospin
chemical potential, which suggests that the pion superfluidity phase
is favored in $\mu_I>64 \text{MeV}$ region.  From the
$\mu_I$-dependent behavior of both pressure difference $\delta{P}$
and the two order parameters, pion condensate and pion mixed
quark-gluon condensate, one can judge the phase transition from
normal phase to pion superfluidity phase is second order. This
conclusion is consistent with the result obtained from lattice
simulation and other model studies.
\begin{figure}[ht]
 \centering {\includegraphics{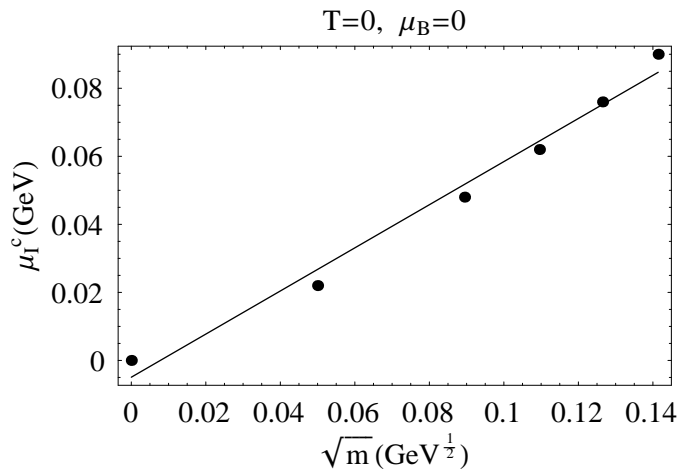}}
  \caption{The dependence of the critical isospin potential $\mu_I^c$ on the current
quark mass from MN model with $\eta=1.06 \text{GeV}$. The current
quark mass ranges from $0$ to $20 \text{MeV}$.}
    \label{F5}
\end{figure}
\par
Note that at zero $\mu$, the scalar functions $A(B,C)_u$ in $S_{uu}$
and $A(B,C)_d$ in $S_{dd}$ at the same point $(\vec{p},w_u,w_d)$ are
complex conjugates, therefore the relation
$\langle{\bar{u}u}\rangle$=$\langle{\bar{d}d}\rangle$ always holds.
Fig. 3 shows that the magnitude of quark condensate monotonically
increases with the increase of $\mu_I$ in the normal phase region,
which is consistent with the dependence of the chiral condensate on
the baryon chemical potential obtained within DSE formalism
\cite{r160}. In the pion superfluidity phase, the magnitude of quark
condensate monotonically decreases with the increase of $\mu_I$,
which is an anticipated result due to the monotonically increasing
behavior of pion condensate with respect to $\mu_I$. This behavior
is similar to the dependence of the chiral condensate on the baryon
chemical potential for the appearance of diquark condensate
\cite{r160}. It is expected that the traditional mixed quark-gluon
condensates, $g\langle\bar{u}\sigma{G}u\rangle$ and
$g\langle\bar{d}\sigma{G}d\rangle$, have the similar
$\mu_I$-dependent behavior. Fig.3 manifests a competitive
relationship between the induced condensates and their corresponding
traditional partners.
\par
In Fig. 4, we display the numerical result of the isospin chemical
potential dependence of the ratio of the pion mixed quark-gluon
condensate to pion condensate. Such a value ranges from
$2.2(\text{GeV})^2$ to $1.7(\text{GeV})^2$, which is also close to
the ratio of the traditional mixed quark-gluon condensate to quark
condensate obtained in the vacuum\cite{r152}. The large magnitude of
the ratio suggests that the induced mixed quark-gluon condensate is
an important parameter within the QCD sum rules at finite $\mu_I$.
In addition,  in contrast with the chiral limit case, the nonzero
pion condensate and nonzero quark condensate suggests that the new
four-quark condensates also have nonzero value in the superfluidity
phase, even at the mean field level.
\par
The critical chemical potential $64\text{MeV}$ is relatively small
in contrast with the vacuum pion mass. As mentioned above, the main
reason for this discrepancy arises from the fact that only
$\gamma_5$ structure is considered in the off-diagram part of the
inverse quark propagator, while the MN model doesn't support the
pseudoscalar bound state beyond the chiral limit when only
$\gamma_5$ structure is contained in the pseudoscalar meson
Bethe-Salpeter amplitude. One can expects that this discrepancy will
become small when we either adopt the other improved effective gluon
propagators in the calculation or include other allowed Dirac
structures such as $\gamma_5\vec{\gamma}\cdot\vec{p}$ in the
off-diagonal part of the inverse quark propagator within the MN
model.
\par
The dependence of the critical isospin potential $\mu_I^c$ on the
current quark mass is plotted in Fig. 5. It is shown that $\mu_I^c$
monotonically increases with the value of current quark mass. Though
$\mu_I^c$ obtained in MN model with only $\gamma_5$ structure
considered is markedly smaller than $m_\pi$, the critical point is
still roughly proportional to the square root of current quark mass
in the range (0-20) MeV.

\vspace{2pt}
\noindent{\textbf{\Large{5. Summary and Remarks}}}
\vspace{2pt}
\par
Using a pedagogical confining model within the framework of GCM, we
have qualitatively investigated the phase structure and condensates
of two-flavor QCD at finite isospin density with zero temperature
and baryon chemical potential. By solving the quark gap equation
through the DSE formalism, we obtained that the truncated DSE type
model supports the pion superfluidity phase transition at high
enough isospin chemical potential. In contrast with the previous
model studies, the obtained gaps responsible for both the chiral
condensate and pion condensate  are all momentum dependent within
the DSE formalism, which are more close to the real world. In
addition, some new types of low-dimensional condensates of QCD
induced by finite isospin chemical potential, such as pion mixed
quark-gluon condensate and mixed four quark condensate,  are
proposed and investigated in this paper.

\par

In the chiral limit with finite isospin chemical potential, the
normal phase is unfavored and the pion superfluidity phase is the
stable ground state. For the case with finite current quark mass,
only the solution corresponding to the normal phase is found in the
gap equations in the region $\mu_I<\mu_I^c$; while for the region
$\mu_I>\mu_I^c$, the normal phase is unfavored and the pion
superfluidity phase is the stable ground state. The distinctly
different phase structure between the chiral limit and the finite
current quark mass suggests that the value of the critical isospin
chemical has close relation with the pion mass. Even though for the
simplicity of numerical study,  the obtained critical point
$\mu_I^c$ in this paper
 is not exactly the pion mass for the case beyond chiral limit, we point out that
the improved calculation (more involved) within the DSE formalism
should confirm $\mu_I^c=m_\pi$.
\par
Furthermore, our calculation shows that in the chiral limit with
finite isospin chemical potential, both quark condensate and
traditional mixed
 quark-gluon condensate vanish in OPE with the appearance of isospin density induced
 pion condensate and pion mixed quark-gluon condensate.
 In the real world, for $\mu_I<\mu_I^c$, quark condensate and mixed quark-gluon condensate
 exist in OPE with the vanishing of pion and pion mixed quark-gluon condensate; for $\mu_I>\mu_I^c$,
 the magnitudes of both isospin density induced condensates
 increase with the increasing of isospin chemical potential, while the magnitude
 of quark condensate decreases with the increasing of
 isospin chemical potential (It is expected that the mixed quark-gluon condensate has the similar behavior).
 Meanwhile, numerical calculations suggest that the induced pion condensate and pion mixed quark-gluon
condensate have the same signs with their corresponding traditional
chiral condensates. We also obtained that the magnitude of the ratio
of pion mixed quark-gluon condensate to pion condensate is close to
the one of traditional quark-gluon condensate to quark condensate in
the vacuum, both for the chiral limit and  for the real world in the
pion superfluidity phase.

\par

Since there is no Fermion sign problem at finite isospin chemical
potential with zero baryon chemical  potential, in principle, the
evaluation of the induced mixed quark-gluon condensate and
four-quark condensate can be investigated through the lattice Monte
Carlo method. The effect of these isospin chemical potential
 induced condensates on the hadron properties can be investigated in the framework of
 QCD sum rules. However, such effects on the hadron properties can more directly
 be explored using suitable Bethe-Salpeter equations in conjunction with the solutions of
 the quark gap equation. A natural extension of the present work is to investigate two
 flavor QCD phase diagram, condensates and hadron properties  at finite isospin chemical potential
 for both nonzero temperature and baryon chemical potential.
\par
\vspace{5pt}
\noindent{\textbf{\Large{Acknowledgements}}}\vspace{5pt}\\
 This work was supported by National Natural Science
Foundation of China under contract 10425521, 10305030 and
10575004, the key Grant Project of Chinese Ministry of Education
(CMOE) under contract No.305001 and the Research Fund for the
Doctoral Program of Higher Education of China under Grant
No.20040001010. One of the author (LYX) thanks also the support of
the Foundation for University Key Teacher by the CMOE. .

\end{document}